\documentclass{article}

\usepackage{microtype}
\usepackage{graphicx}
\usepackage{subfigure}
\usepackage{booktabs} 

\usepackage{hyperref}

\usepackage[accepted]{arxiv}
\usepackage{longtable}
\usepackage{listings}
\usepackage{xcolor}
\definecolor{codegreen}{rgb}{0,0.6,0}
\definecolor{codegray}{rgb}{0.5,0.5,0.5}
\definecolor{codepurple}{rgb}{0.58,0,0.82}
\definecolor{backcolour}{rgb}{0.95,0.95,0.92}

\lstdefinestyle{mystyle}{
    backgroundcolor=\color{backcolour},   
    commentstyle=\color{codegreen},
    keywordstyle=\color{magenta},
    numberstyle=\tiny\color{codegray},
    stringstyle=\color{codepurple},
    basicstyle=\ttfamily\footnotesize,
    breakatwhitespace=false,         
    breaklines=true,                 
    captionpos=b,                    
    keepspaces=true,                 
    numbers=none,                    
    numbersep=5pt,                  
    showspaces=false,                
    showstringspaces=false,
    showtabs=false,                  
    tabsize=2,
    rulesep=0pt,
    framextopmargin=0pt,
}
\usepackage{amsmath}
\usepackage{amssymb}
\usepackage{mathtools}
\usepackage{amsthm}
\usepackage{pifont}

\usepackage[capitalize,noabbrev]{cleveref}

\usepackage{array}
\newcolumntype{L}[1]{>{\raggedright\let\newline\\\arraybackslash\hspace{0pt}}m{#1}}
\newcolumntype{C}[1]{>{\centering\arraybackslash}p{#1}}
\newcolumntype{R}[1]{>{\raggedleft\let\newline\\\arraybackslash\hspace{0pt}}m{#1}}
 \usepackage{multirow}

\theoremstyle{plain}

\theoremstyle{definition}

\theoremstyle{remark}

\usepackage[textsize=tiny]{todonotes}
\usepackage{algorithm}
\usepackage{algpseudocode}
\usepackage{fixltx2e}

\icmltitlerunning{Improving Code Generation by Training with Natural Language Feedback}

\begin{document}

\twocolumn[
\icmltitle{Improving Code Generation by Training with Natural Language Feedback}



\icmlsetsymbol{equal}{*}

\begin{icmlauthorlist}
\icmlauthor{Angelica Chen}{nyu}
\icmlauthor{Jérémy Scheurer}{nyu,far}
\icmlauthor{Tomasz Korbak}{nyu,far,su}
\icmlauthor{Jon Ander Campos}{nyu,ba}
\icmlauthor{Jun Shern Chan}{nyu,far}
\icmlauthor{Samuel R. Bowman}{nyu}
\icmlauthor{Kyunghyun Cho}{nyu,gen,cif}
\icmlauthor{Ethan Perez}{nyu,far,ant}
\end{icmlauthorlist}

\icmlaffiliation{nyu}{New York University}
\icmlaffiliation{far}{FAR AI}
\icmlaffiliation{ba}{HiTZ Center, University of the Basque Country UPV/EHU}
\icmlaffiliation{su}{University of Sussex}
\icmlaffiliation{gen}{Genentech}
\icmlaffiliation{cif}{CIFAR LMB}
\icmlaffiliation{ant}{Anthropic}

\icmlcorrespondingauthor{Angelica Chen}{angelica.chen@nyu.edu}
\icmlcorrespondingauthor{Ethan Perez}{ethan@anthropic.com}
\icmlkeywords{Machine Learning, ICML}

\vskip 0.3in
]



\printAffiliationsAndNotice{}  


\begin{abstract}
The potential for pre-trained large language models (LLMs) to use natural language feedback at inference time has been an exciting recent development. We build upon this observation by formalizing an algorithm for learning from natural language feedback at training time instead, which we call Imitation learning from Language Feedback (ILF). ILF requires only a small amount of human-written feedback during training and does not require the same feedback at test time, making it both user-friendly and sample-efficient. We further show that ILF can be seen as a form of minimizing the KL divergence to the ground truth distribution and demonstrate a proof-of-concept on a neural program synthesis task. We use ILF to improve a \textsc{Codegen-Mono 6.1B} model's pass@1 rate by 38\% relative (and 10\% absolute) on the Mostly Basic Python Problems (MBPP) benchmark, outperforming both fine-tuning on MBPP and fine-tuning on repaired programs written by humans. Overall, our results suggest that learning from human-written natural language feedback is both more effective and sample-efficient than training exclusively on demonstrations for improving an LLM's performance on code generation tasks.
\end{abstract}

\begin{figure}[ht!]
    \centering
    \includegraphics[width=\columnwidth]{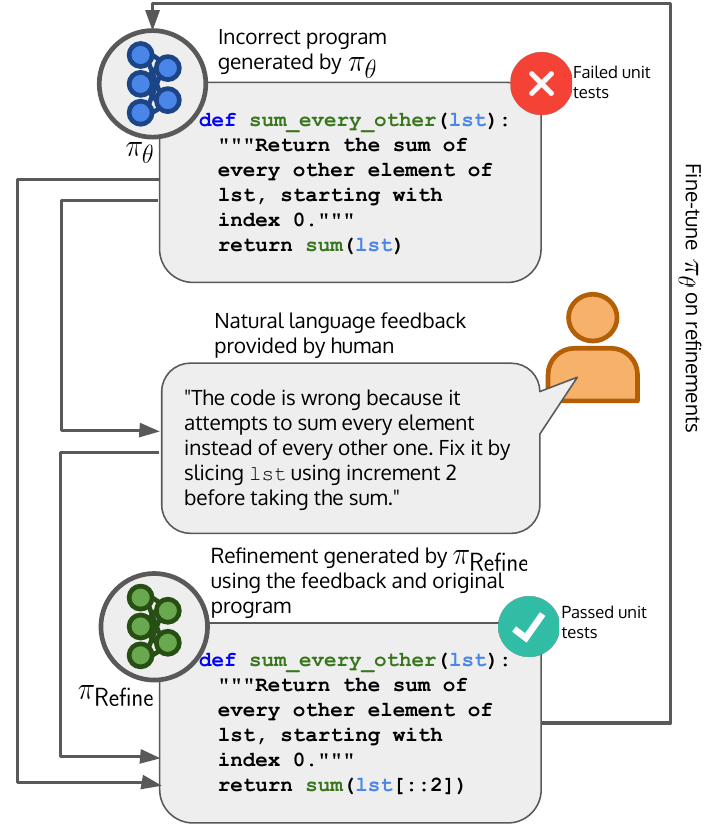}
    \caption{An overview of imitation learning from language feedback (ILF) for code generation. Given an initial LLM $\pi_{\theta}$, we sample programs from $\pi_{\theta}$ that do not pass unit tests (indicated by the red X). Human annotators write natural language feedback for the incorrect program and a model $\pi_\text{Refine}$ generates a \emph{refinement} - \emph{i.e.} an improved version of the original program that incorporates the feedback and passes the unit tests (indicated by the green checkmark). Finally, we fine-tune $\pi_{\theta}$ on the refinements.}
    \label{fig:diagram}
\end{figure}

\section{Introduction}
An important task for the field of software engineering is program synthesis, the automatic generation of computer programs from an input specification (\emph{e.g.} a natural language task description or a set of input-output examples) \citep{manna1971prog_synth}. Effective program synthesis can not only improve the efficiency of software developers \citep{code_completion_productivity}, but also increase the accessibility of writing code in general.
Recently, pre-trained large language models (LLMs) have demonstrated impressive success on program synthesis \citep[\textit{inter alia}]{chen2021codex, li2022alphacode, austin2021program, Nijkamp2022CG, xu2022evaluation} but still struggle to consistently generate correct code, even with large-scale pre-training \citep{chen2021codex}.

We hypothesize that these failures can be largely attributed to modern LLM pre-training set-ups. For instance, code pre-training datasets consist mostly of unfiltered code scraped from the Internet, which contains a significant number of security vulnerabilities \citep{kang2022llm_bugs} and bugs \citep{chen2021codex}.
This training signal also consists exclusively of offline demonstrations, without any signal from trial-and-error or interactive guidance that penalizes the model's buggy outputs. As such, we hypothesize that supervising LLMs with explicit human-written feedback on the model's own outputs can be more effective at training models to produce functionally correct code.

In particular, an intuitive and rich form of feedback to provide to LLMs is \textit{natural language} feedback.
We argue that LLMs are naturally able to incorporate written feedback, which has been shown to significantly improve a code generation model's pass rates when the feedback is provided at test time \citep{Nijkamp2022CG, austin2021program}. In our work, we build upon this observation by exploring the use of natural language feedback during the training process itself, rather than just during inference. We conjecture that such feedback provides expressive and targeted information about a code generation model's current failings in a sample-efficient manner. More broadly, this approach also represents a weak version of \emph{scalable oversight} \citep{Bowman2022MeasuringPO}, in that model overseers can improve a model merely by evaluating its outputs, without manually generating new demonstrations, in a way that takes advantage of the capabilities that are being supervised.

To train LLMs with language feedback, we propose an algorithm called Imitation learning from Language Feedback (ILF; Algorithm \ref{alg:il}), which extends the work of \citet{scheurer2022training}, who study the impact of learning from language feedback on text summarization models. \citet{scheurer2022training} improves a summarization model by training the base model on improved summaries generated from the model's original summaries and human-written feedback. Our work builds upon \citet{scheurer2022training} in a number of ways: (1) by formalizing the algorithm and generalizing it into a form that can be applied to any task (our ILF algorithm in Section \ref{sec:ilf-method}), (2) by detailing how the reward function can be adapted for code generation, and (3) by demonstrating a proof-of-concept of ILF for code generation.\footnote{We open-source our code and annotated data at \url{https://github.com/nyu-mll/ILF-for-code-generation}.}
ILF improves the correctness of programs generated by a baseline code generation model $\pi_{\theta}$ by training a separate model $\pi_\text{Refine}$ to use language feedback to repair the incorrect $\pi_{\theta}$-generated programs. (We refer to the repaired programs as \emph{refinements}.) We then improve $\pi_{\theta}$ by fine-tuning it on the $\pi_\text{Refine}$-generated refinements that pass unit tests, yielding a final improved model $\pi_{\theta^*}$. This procedure may be run iteratively to continue improving the model, which we show can be seen as minimizing the expected KL divergence from a target ground truth distribution (Section \ref{sec:method-overall}).

We demonstrate a proof of concept of ILF for code generation by showing that it improves a \textsc{CodeGen-Mono 6.1B} model's pass@1 rate on the Mostly Basic Python Problems (MBPP) benchmark \citep{odena2021mbpp} by 38\% relative (10\% absolute) over its zero-shot performance. It also outperforms fine-tuning on the MBPP-provided code by 64\% (14\% absolute, see Section \ref{sec:ilf-results}).
We further find that the refinements generated during ILF do indeed leverage the human-written feedback (Section \ref{sec:pi_ref_few_shot}) -- when the feedback is unhelpful or irrelevant, we observe steep drops in code correctness. The quality of the feedback is also crucial -- LLM-generated feedback yields far lower final pass rates than human-written feedback (Section \ref{sec:llm-feedback}).
Despite the success of our approach, we still observe concrete limitations -- for instance, $\pi_\text{Refine}$ is less effective at incorporating feedback when the feedback addresses multiple bugs (Section \ref{sec:pi-refine-feedback-num-bugs}), which suggests headroom for future work or more capable LLMs to base $\pi_\text{Refine}$ on.
Overall, our results -- as well as our additional results on text summarization, using a similar technique in \citet{scheurer2023training} -- suggest that human-written feedback is a powerful, information-rich form of supervision for LLMs.

\section{Method}\label{sec:method-overall}
\begin{algorithm*}[th!]
   \caption{Imitation learning from natural language feedback for code generation.}
   \label{alg:il}
\begin{algorithmic}[1]
   \State {\bfseries Input:} Dataset $\mathcal{D}$, initial LLM $\pi_{\theta}$, unit test verification function $\textsc{Eval}$, LLM $\pi_\text{Refine}:\mathcal{V}^*\to [0,1]$ trained to incorporate feedback into code
   \State $C \leftarrow \{(x_0,t,u) \, | \, x_0\sim \pi_{\theta_k}(\cdot | t), \textsc{Eval}(x_0,t)=0, (t,u)\in\mathcal{D}\}$ \label{alg-step:init-code-sample}
   \State $C_\text{annotated}\leftarrow \{(x_0,f,t)\,|\,(x_0, t, u) \in C\}$ \algorithmiccomment{Humans write feedback $f$ for $x_0\in C$.} \label{alg-step:feedback}
   \State $R\leftarrow \{(t,x_1)\sim \pi_\text{Refine}(\cdot \,|\, t,x_0,f)\,|\, \textsc{Eval}(x_1,t)=1, (x_0, f, t) \in C_\text{annotated} \}$  \algorithmiccomment{$\pi_\text{Refine}$ generates refinements $x_1$ that incorporate feedback $f$ into $x_0$.}
   \State $\pi_{\theta^*} \leftarrow \textsc{Finetune}(\pi_{\theta}, R)$ 
\end{algorithmic}
\end{algorithm*}

\subsection{Preliminaries}
Here, we formally describe the problem we aim to tackle, before introducing our algorithm.
Suppose we start with vocabulary $\mathcal{V}$ and a pre-trained language model $\pi_\theta$ parameterized by $\theta$. $\pi_\theta: \mathcal{V^*}\to[0,1]$ is a probability distribution over sequences of tokens $x \in \mathcal{V}^*$, where $\mathcal{V}^*$ is the Kleene closure of $\mathcal{V}$. We also have a dataset of tasks $\mathcal{D}=\{(t, u)\}$. A task $(t,u)$ consists of a task description $t \in \mathcal{T}$ (\emph{e.g.} ``Write a function that computes the prime factorization of an input integer.") and a suite $ u =\textsc{UnitTests}(t) \in \mathcal{U}$ of unit tests associated with task $t$. Finally, let $\textsc{Eval}:\mathcal{V}^*\times\mathcal{T}\to \{0,1\}$ be a unit test verification function that indicates whether a program $x\sim \pi_\theta(\cdot \,|\, t)$ passes all the unit tests in $\textsc{UnitTests}(t)$:
\begin{equation}\label{func:unit-test-verif}
    \textsc{Eval}(x,t) \coloneqq \left\{
	\begin{array}{ll}
		1,  & \mbox{if } x \text{ passes test suite\ } \textsc{UnitTests}(t), \\
		0, & \mbox{otherwise} 
	\end{array}
\right.
\end{equation}
We also define a fine-tuning function $\textsc{Finetune}(\pi_\theta,\mathcal{D})$ that applies a gradient-based optimization algorithm to $\pi_\theta$ using the associated loss objective calculated over dataset $\mathcal{D}$.
\subsection{Imitation Learning From Language Feedback}
\label{sec:ilf-method}
Our goal is to sample a diverse set of high-quality programs $x_1\sim \pi_\theta(\cdot | t)$ for any given task $t$ sampled from the task distribution $p(t)$. We do so by fitting an auto-regressive LLM $\pi_\theta$ to approximate a ground truth distribution $\pi_t^*(x_1)$ that assigns a probability to $x_1$ that is proportional to its quality, as measured by a reward function $R$. Fitting $\pi_\theta$ to approximate $\pi_t^*$ can be seen as minimizing the expected KL divergence from $\pi_t^*$ to $\pi_\theta$ over the task distribution $p(t)$:
\begin{equation} \label{eqn:kl-obj}
    \min_\theta \underset{t\sim p(t)}{\mathbb{E}} \left[ \mathrm{KL} (\pi_t^*, \pi_\theta(\cdot\,|\,t) ) \right]
\end{equation}
where
\begin{equation}
    \pi_t^*(x_1)\propto \exp\left(\beta R(x_1, t)\right)
\end{equation}
In this work we use the unit test verification function $\textsc{Eval}$ directly as our reward function $R$, but $R$ can also be a function of any number of other signals, such as stack traces or compiler outputs.

Minimizing the objective in Equation \ref{eqn:kl-obj} is equivalent to supervised learning, \emph{i.e.} minimizing the cross-entropy loss:
\begin{equation}
    \mathcal{L}(\theta) = -\underset{t\sim p(t)}{\mathbb{E}} \left[ \mathcal{L}_\theta(t)\right],
\end{equation}
where
\begin{equation}
    \mathcal{L}_\theta(t) = \sum_{x_1} \pi_t^* (x_1) \log \pi_\theta (x_1|t).
\end{equation}
Rather than computing this loss over the exponentially large space of all possible $x_1$'s, we instead use Monte-Carlo sampling over a small set of $x_1$'s drawn from $\pi_t^*$. However, this is still intractable because we cannot sample directly from $\pi_t^*$. Instead, we approximate $\pi_t^*$ using importance sampling with a proposal distribution $q_t(x_1)$:
\begin{equation}
    \mathcal{L}_\theta(t) = \sum_{x_1} q_t(x_1) \frac{\pi_t^* (x_1)}{q_t(x_1)} \log \pi_\theta (x_1|t)
\end{equation}which assigns higher weights to higher quality programs $x_1$.

\subsection{Proposal Distribution $q$}
\label{sec:opt-q}

Intuitively, we aim to design $q_t$ to be as close as possible to $\pi_t^*$, which we accomplish by incorporating pieces of natural language feedback $f$ that give information about how to transform a low-reward program $x_0$ into a higher-reward program $x_1$. This can be achieved by (i) identifying a program $x_0\sim \pi_\theta(\cdot|t)$ that does not currently pass the test suite (\emph{i.e.} $\textsc{Eval}(x_0,t)=0$), (ii) asking for natural language feedback $f$ about bugs in $x_0$, (iii) using $f$ to transform the original program $x_0$ into a \emph{refinement} $x_1$ that incorporates the feedback and passes the test suite (\emph{i.e.} $\textsc{Eval}(x_1,t)=1$), and (iv) assigning higher weight to $x_1$. 

We can formalize this procedure as follows. Let $\pi_\psi(x_1|t, x_0, f)$ be a distribution over programs $x_1$ that improve $x_0$ by incorporating the feedback $f$ and $p_\mathcal{F}(f\,|\, t, x_0, \textsc{Eval}(x_0, t)=0)$ be the distribution of pieces of feedback $f$ for incorrect program $x_0$ and task $t$. We can then define our proposal distribution as:
\begin{align}
q_t(x_1) = 
      \sum_{x_0, f} \ &\pi_\theta(x_0|t) \times \delta_0\left(\textsc{Eval}(x_0,t)\,|\, x_0, t)\right)\nonumber\\ 
      &\times p_\mathcal{F}(f|t,x_0, \textsc{Eval}(x_0, t)=0) \nonumber \\
      &\times \pi_\psi(x_1|t,x_0,f)  \nonumber\\ 
      &\times  \delta_1(\textsc{Eval}(x_1,t)\,|\, t, x_1), \label{eqn:q-factorization}
\end{align}
where $\delta_0$ and $\delta_1$ are the Dirac delta distributions centered at 0 and 1, respectively.
Then this proposal distribution is guaranteed to place higher probability mass on higher-quality programs (in terms of unit test pass rate) than $\pi_\theta$ since the term $\delta_1(\textsc{Eval}(x_1,t)\,|\, t, x_1)$ equals 0 for incorrect programs $x_1$.

We approximate sampling from $q$ by considering each of the terms in Equation \ref{eqn:q-factorization} in order:
\begin{enumerate}
    \item \label{em-sampling-first} We first sample from $\pi_\theta(x_0|t) \times \delta_0\left(\textsc{Eval}(x_0,t)\,|\, x_0, t)\right)$ by rejection sampling from $\pi_{\theta}$. In other words, we sample programs $x_0$ from $\pi_\theta$ for task $t$ and only keep those  that fail the test suite (\emph{i.e.} $\textsc{Eval}(x_0,t)=0$; step 2 of Algorithm \ref{alg:il}).
    \item We approximate sampling from $p_\mathcal{F}(f|t,x_0, \textsc{Eval}(x_0, t)=0)$ by having humans annotate programs $x_0$ (paired with their corresponding task descriptions $t$ and test suites $u$) with natural language feedback (step 3 of Algorithm \ref{alg:il}). 
    \item \label{em-sampling-mref} We approximate sampling from $\pi_\psi(x_1|t,x_0,f)$ by sampling from $\pi_\text{Refine}$, a model capable of generating refinements given the task description, original programs, and human-written feedback.
    \item Finally, the term $\delta_1(\textsc{Eval}(x_1,t)\,|\, t, x_1)$ corresponds to another filter: we only keep refined programs $x_1$ that pass the test suite. \label{em-sampling-last}
\end{enumerate}

Next, we consider more concrete details of how this sampling is accomplished.

\paragraph{Training $\pi_\text{Refine}$}
ILF assumes the availability of feedback but not necessarily of the repaired code/refinements, for a variety of reasons. We assume that program synthesis may be a task for which writing high-level natural language feedback is often less laborious than performing program repair. Although writing feedback involves identifying at a high level what is wrong with the program and how it should be fixed, program repair may involve the additional steps of refactoring, looking through documentation, and testing. Moreover, past work \citep{austin2021program, Nijkamp2022CG} has indicated that certain large LLMs can proficiently incorporate the feedback at inference time, assuming access to accurate and high-quality feedback. As such, ILF assumes access to some model $\pi_\text{Refine}$ that is capable of producing a refinement given the original program and feedback.

$\pi_\text{Refine}$ can take a variety of forms, but we fine-tune a pre-trained \textsc{CodeGen-Mono 6.1B} model as our $\pi_\text{Refine}$. We create a training dataset for $\pi_\text{Refine}$ by further annotating a subset of $C_\text{annotated}$ with refinements $x_1$ that repair incorrect programs $x_0$ by incorporating feedback $f$, such that $\textsc{Eval}(x_1,t)=1$ for $(x_0,f,t)\in C_\text{annotated}$. Further details of our dataset and annotation procedure are in Section \ref{sec:experiments-and-results}.

\begin{figure*}[ht!]
\begin{center}
\centerline{\includegraphics[width=0.8\textwidth]{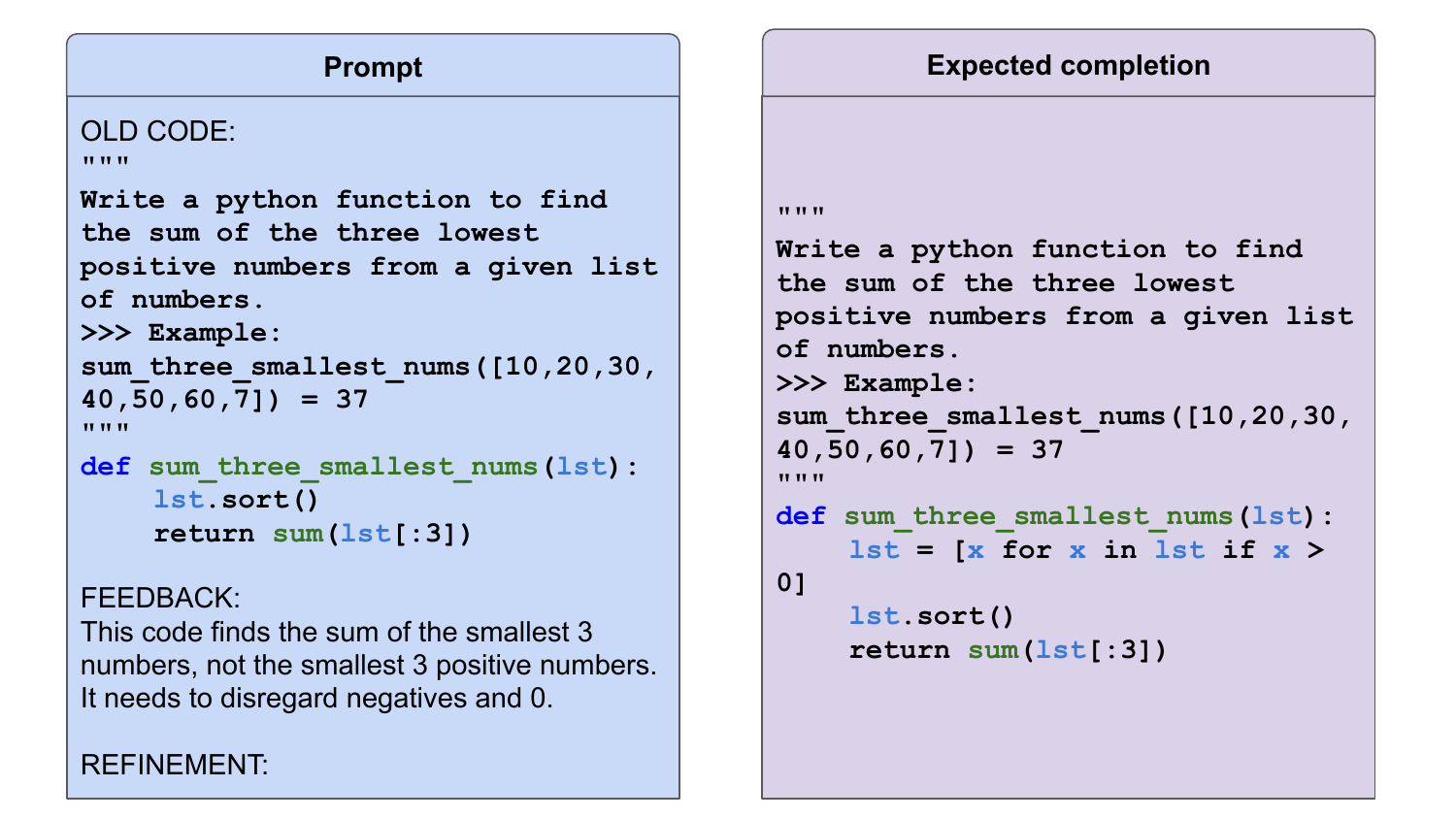}}
\caption{An example of a zero-shot LLM prompt for repairing incorrect code based on human-written feedback.}
\label{fig:feedback-prompt}
\end{center}
\end{figure*}

\section{Experiments and Results}
\label{sec:experiments-and-results}

Having described our high-level approach, we now explain the experimental setup we use to test ILF.

\paragraph{Dataset} 
We train and evaluate our models on the Mostly Basic Python Problems (MBPP) dataset \citep{odena2021mbpp}. MBPP contains 974 Python programming tasks designed to be solvable by entry-level coders. Each task contains a natural language task description $t$ (\emph{e.g.}, ``Write a function to return the prime factorization of the input.''), a gold solution, and a suite $u$ of three unit tests. Since the task descriptions are sometimes ambiguous, we include one unit test in the task description. The addition of the unit test helps to specify the input and output format of each task.  We hold out the remaining unit tests for the evaluation of our generated programs.

MBPP includes a designated prompt/training/validation/test split of the dataset, but we re-split the dataset into the following splits:
\begin{itemize}
    \item MBPP\textsubscript{Refine}: These are tasks with IDs in the range 111-310 for which \textsc{CodeGen-Mono 6.1B} did not generate any correct completions. This split is used to train $\pi_\text{Refine}$.
    \item MBPP\textsubscript{Train}: These are tasks with IDs in the range 311-974 for which \textsc{Codegen-Mono 6.1B} did not generate any correct completions. This split is first used to evaluate the correctness of refinements generated by $\pi_\text{Refine}$. Then, the correct refinements in this split are used to train $\pi_\theta$ to obtain $\pi_{\theta^*}$ (step 5 in Algorithm \ref{alg:il}).
    \item MBPP\textsubscript{Test}: These are tasks with IDs in the range 11-110 that we use to evaluate the final performance of $\pi_{\theta^*}$. Unlike the previous two splits, we use \emph{all} tasks in this split, rather than only the tasks for which \textsc{CodeGen-Mono 6.1B} did not originally generate correct programs for. This allows us to better compare the baseline performance of $\pi_\theta$ with that of $\pi_{\theta^*}$.
 \end{itemize}
We use this modified split so that a larger portion of the dataset can be used to train the final model $\pi_{\theta^*}$, whereas smaller portions are allocated for training $\pi_\text{Refine}$ and evaluating $\pi_{\theta^*}$. We do not make use of the prompt split (IDs 1-10).

\paragraph{Models}

Throughout this paper, we use a pre-trained \textsc{CodeGen-Mono 6.1B} model \citep{Nijkamp2022CG} as our $\pi_{\theta}$. It is pre-trained sequentially on \textsc{ThePile} \citep{gao2020pile}, \textsc{BigQuery} \citep{Nijkamp2022CG}, and \textsc{BigPython} \citep{Nijkamp2022CG}. We selected this model because it is open-source, can be fine-tuned on a single $4\times 100$ A100 (80 GB) node, and demonstrated pass@k scores comparable to \textsc{Codex-12B} \citep{chen2021codex, Nijkamp2022CG}.

To implement our algorithm, we independently fine-tune two separate instances of \textsc{CodeGen-Mono 6.1B} to create $\pi_\text{Refine}$ and the final model $\pi_{\theta^*}$. We train $\pi_\text{Refine}$ using pairs of incorrect programs and human-written feedback as inputs, with human-written refinements as targets (using the format in Figure \ref{fig:feedback-prompt}). In contrast, we train $\pi_{\theta^*}$ using natural language task descriptions from MBPP as the inputs and $\pi_\text{Refine}$-generated refinements as the targets. Further training details are in Appendix \ref{appendix:training-details}.

\paragraph{Evaluation}
We evaluate all code generations in this paper using the \emph{pass@k} metric introduced in \citet{kulal2019spoc}. It estimates the rate for which $\geq$1 of $k$ model samples passes all the unit tests. We use the empirical estimate of this quantity from \citet{chen2021codex}, an unbiased estimator given by:
\begin{equation}
\text{pass@k} \coloneqq \mathbb{E}_\text{task} \left[1-\frac{\binom{n-c}{k}}{\binom{n}{k}}\right]
\end{equation}
for $n$ total programs (where $n\geq k$) and $c$ correct programs for the given task.

\paragraph{Human Annotation}
\label{sec:surge}
We hire annotators via Surge AI\footnote{\url{www.surgehq.ai}} to write both natural language feedback and refinements for incorrect programs generated by $\textsc{CodeGen-Mono 6.1B}$. For each task that \textsc{CodeGen-Mono 6.1B} generated no correct programs for, we ask the workers to first select one of the incorrect programs to write feedback and refinement for. We specify that the workers should select a sample that seems relatively easy to correct (\emph{i.e.} could be minimally corrected to pass the unit tests). Then, they are asked to write feedback that describes what is wrong with the current code and how to fix it. For the refinement, they are asked to copy over the original code and make the \emph{minimum number of edits necessary} to incorporate the feedback and pass all the unit tests. The full set of worker instructions can be found in Appendix \ref{appendix:annotator-instructions}.

We keep all annotations for which the refinement passes all tests in the task's test suite, the feedback is correct (as manually verified by the authors), and the Levenshtein edit distance between the refinement and the original program is less than 50\% of $\max(\mathrm{len}(\text{refinement}), \mathrm{len}(\text{original program}))$. The final dataset consists of 195 triples of (incorrect program, human-written feedback, human-written refinement).
On average, workers are paid \$23 per annotated sample and take 27 minutes/sample, with a 10th percentile of 4 minutes and a 90th percentile of 43 minutes.

Although the ILF algorithm only requires the collection of human-written feedback for the tasks in MBPP\textsubscript{Train} (assuming access to some $\pi_\text{Refine}$ that is already fine-tuned or can generate refinements via few-shot prompting), we collect both human-written feedback and refinement for all splits of the data so that we can conduct further analyses of our method. For instance, this allows us to compare fine-tuning on $\pi_\text{Refine}$-generated refinements with fine-tuning on human-written refinements. When scaled to other pairs of model and task, ILF requires new feedback annotations, but it is possible that using ILF on one dataset will improve the model's abilities on another dataset for a similar task. We leave analyses of scaling ILF across different tasks and models to future work.

\begin{table}[th!]
\caption{Initial zero-shot \textsc{CodeGen-Mono 6.1B} performance on the entire MBPP dataset. ``1+ Correct" refers to the percentage of tasks for which \textsc{CodeGen-Mono 6.1B} generated at least one program that passed all unit tests.}
\label{tab:codegen-init-mbpp}
\begin{center}
\small
\begin{tabular}{p{1.5cm}R{4.5cm}} \toprule
Metric & Zero-Shot \textsc{CodeGen-Mono 6.1B}\\
\midrule
Pass@1 & 31\% \\
Pass@10 & 63\% \\
1+ Correct & 67\% \\
\bottomrule
\end{tabular}
\end{center}
\end{table}

\begin{table}[th!]
\caption{Evaluations of 1-shot refinements generated by \textsc{CodeGen-Mono 6.1B} (before ILF) given either related or unrelated text feedback in the prompt. Feedback is provided only for tasks on which \textsc{CodeGen-Mono 6.1B} previously did not output any correct programs.}
\label{tab:1-shot-incorp-feedback}
\begin{center}
\begin{small}
\begin{tabular}{p{0.5\columnwidth}rr}
\toprule
\multirow{2}{=}{Prompt Type} & \multicolumn{2}{c}{\textsc{CodeGen-Mono 6.1B}} \\
 & Pass@1 $\uparrow$ & Pass@10 $\uparrow$  \\ 
  \midrule
 Code + feedback & 2.0\% &  13.8\%   \\  
 Code + unrelated feedback & 0.4\% & 4.0\%  \\ 
\bottomrule
\end{tabular}
\end{small}
\end{center}
\end{table}

\subsection{\textsc{CodeGen-Mono 6.1B} Incorporates Feedback}\label{sec:pi_ref_few_shot}

We first verify that our baseline model can use feedback to repair incorrect code, a pre-requisite for ILF to work.
We evaluate $\textsc{CodeGen-Mono 6.1B}$'s ability to generate refinements given pairs of (incorrect code, natural language feedback), both in a few-shot manner and after fine-tuning. Feedback is only required for tasks for which $\pi_{\theta}$ is initially unable to produce a correct response, so we first evaluate $\textsc{CodeGen-Mono 6.1B}$ zero-shot on all of MBPP, generating 30 programs per task with temperature 0.8. Table \ref{tab:codegen-init-mbpp} shows the resulting pass rates. There were 321 tasks for which zero-shot $\textsc{CodeGen-Mono 6.1B}$ yielded no correct samples (from Table \ref{tab:codegen-init-mbpp}: $(100\%-67\%)
\times 974\text{ tasks}\approx 321$).
We then annotate one incorrect program per task with both feedback and refinement, as described in Section \ref{sec:surge}.

\paragraph{Few-Shot Feedback Incorporation} We use the human feedback annotations to create few-shot feedback prompts, formatted as in Figure \ref{fig:feedback-prompt}. We evaluate \textsc{CodeGen-Mono 6.1B}'s ability to produce refinements that incorporate the feedback and pass the unit tests. However, producing a refinement that passes the unit tests does not guarantee that the feedback has been incorporated; there can be multiple solutions to a programming task, including ones that are functional but completely different and not using the feedback to improve upon the original code. Alternatively, the model may already be able to repair programs without feedback.
Thus, we also evaluate the pass rate after shuffling the feedback samples in the dataset, to evaluate if the model's ability to repair code degrades when presented with unrelated feedback.

The results are shown in Table \ref{tab:1-shot-incorp-feedback}. \textsc{CodeGen-Mono 6.1B}'s ability to incorporate relevant feedback on this particular set of program is low, with pass@10 reaching only 13.8\%. However, the gap in accuracy between \textsc{CodeGen-Mono 6.1B}-generated refinements on relevant versus irrelevant feedback is significant, with pass@10 decreasing by 71\% (relative; 13.8\% $\rightarrow$ 4.0\%), indicating that the model is indeed using the feedback.

\paragraph{Training $\pi_\text{Refine}$}
Next, we examine whether we can improve our ability to repair programs given feedback by fine-tuning a separate model specifically to perform this task.
Our training examples consist of triples of incorrect program, human-written feedback, and human-written refinement. We train the model to maximize the likelihood of the refinement given the program and feedback. The incorrect programs were generated by \textsc{CodeGen-Mono 6.1B} zero-shot on MBPP tasks, and the feedback and refinements were written by human annotators, as discussed in Section \ref{sec:experiments-and-results}. We only included tasks for which none of \textsc{CodeGen-Mono 6.1B}'s generated programs were correct, yielding 44 tasks in the training dataset (forming the split MBPP\textsubscript{Refine}) and 128 tasks in the evaluation dataset (forming the split MBPP\textsubscript{Train}). We asked human annotators to write refinements of the original code that incorporated their own previously written feedback, passed the unit tests, and made only minimal edits to the code (see Section \ref{sec:surge}). The format of the training data also matched the few-shot prompt format (Figure \ref{fig:feedback-prompt}) but without the in-context examples of refinements. We denote this model as $\pi_\text{Refine}$, as described in Section \ref{sec:opt-q}.

\begin{table}[th]
\caption{Pass rates of $\pi_\text{Refine}$-generated refinements versus zero-shot \textsc{CodeGen-Mono 6.1B} programs for tasks in MBPP\textsubscript{Train}.}
\label{tab:mref}
\begin{center}
\small
\begin{tabular}{p{0.2\columnwidth}R{0.1\columnwidth}R{0.55\columnwidth}}
\toprule
 Metric & $\pi_\text{Refine}$ & Zero-shot \textsc{CodeGen-Mono 6.1B} \\
 \midrule
 Pass@1 & 19\% & 0\% \\
 Pass@10 & 47\% & 0\% \\
1+ correct & 61\% & 0\% \\ \bottomrule
\end{tabular}
\end{center}
\vskip -0.1in
\end{table}

Table \ref{tab:mref} shows the pass rates for $\pi_\text{Refine}$ on the evaluation dataset, which were produced by sampling 30 refinements per task with temperature 0.8. Fine-tuning significantly improves $\textsc{CodeGen-Mono 6.1B}$'s ability to incorporate feedback compared to 1-shot refinement, increasing pass rates more than three-fold (2$\rightarrow$19\% pass@1, 13.8$\rightarrow$47\% pass@10, from Tables \ref{tab:1-shot-incorp-feedback} and \ref{tab:mref}). Furthermore, 61\% of tasks had at least one correct refinement. This is particularly significant when considering the fact that we selected only tasks for which a non-finetuned $\textsc{CodeGen-Mono 6.1B}$ model did not originally output any correct programs for (the rightmost column in Table \ref{tab:mref}). For the 61\% of validation tasks that $\pi_\text{Refine}$ generated a correct refinement for, we randomly selected one such correct program for each task to form the training dataset for our final model $\pi_{\theta^*}$, yielding a final training dataset of 78 examples.

\begin{table*}[th!]
\caption{Final performance of $\pi_{\theta^*}$ on MBPP\textsubscript{Test}, compared to other ablations and baselines. All results are calculated using 30 output samples with temperature 0.8. All the methods are built on the \textsc{CodeGen-Mono 6.1B} model.}
\label{tab:final-results}
\begin{center}
\begin{small}
\begin{tabular}{lllrr}
\toprule
Method  & Feedback Source & Fine-Tuning Data & \multicolumn{2}{c}{Pass Rates of $\pi_{\theta^*}$} \\
 &  & & Pass@1 & Pass@10  \\
  \midrule
ILF & Humans & $\pi_\text{Refine}$ Refinements & \textbf{36\%} & \textbf{68\%}	 \\ \midrule
\multirow{2}{*}{Ablations} & 1-shot InstructGPT & 1-shot InstructGPT Refinements  & 19\% & 55\%  \\ 
& 2-shot InstructGPT & 2-shot InstructGPT Refinements & 25\% & 59\%  \\ \midrule
\multirow{2}{*}{Gold Standards} & - & MBPP Gold & 22\% & 63\%  \\
& - & Human Refinements& 33\% & \textbf{68\%} \\ \midrule
\multirow{1}{*}{Baseline (zero-shot)} & - & - & 26\% & 59\%  \\
\bottomrule
\end{tabular}
\end{small}
\end{center}
\vskip -0.1in
\end{table*}

\subsection{ILF Yields Pass Rates Higher Than Fine-Tuning on Gold Data or Human-Written Programs Alone} \label{sec:ilf-results}
Given that our refinements improve over the initial programs, we now fine-tune on the refinements to improve our code generation model.
As discussed earlier, we use the correct refinements (as evaluated by the unit tests) that $\pi_\text{Refine}$ generated for its evaluation dataset as the training dataset for $\pi_{\theta^*}$. Since $\pi_{\theta^*}$ is meant to generate code from a natural language task description (rather than to incorporate feedback into a refinement), the inputs of our training dataset are the MBPP prompts and the targets are the 78 $\pi_\text{Refine}$-generated refinements described in the previous section. We also compare the performance of $\pi_{\theta}^*$ against that of \textsc{CodeGen-Mono 6.1B} evaluated in a zero-shot manner, \textsc{CodeGen-Mono 6.1B} fine-tuned on the gold programs from the MBPP dataset, and \textsc{CodeGen-Mono 6.1B} fine-tuned on our human-written refinements. For all fine-tuning experiments, we train on programs corresponding to the same set of task IDs as the ones used in $\pi_{\theta^*}$'s training dataset.

Additionally, we evaluate the impact of ablating the human annotations in our algorithm by using an LLM in place of humans to generate the feedback and refinements (replacing steps 3 and 4 in Algorithm \ref{alg:il}). For the LLM, we use GPT-3.5 fine-tuned with Feedback Made Easy (FeedME; \texttt{text-davinci-002} on the OpenAI API)\footnote{Details at \href{https://beta.openai.com/docs/model-index-for-researchers}{beta.openai.com/docs/model-index-for-researchers}}. We refer to this model as InstructGPT, which is the series of OpenAI models that FeedME belongs to \citep{openai_mir}. We use InstructGPT to generate both the feedback and refinements on the original programs.
We then fine-tune $\textsc{CodeGen-Mono 6.1B}$ on the model-generated refinements.

The results of our ILF algorithm compared to the baselines and ablations are shown in Table \ref{tab:final-results}. ILF yields the highest pass@1 and pass@10 rates, despite how few samples of feedback and refinements we use. The pass@1 rate in particular shows a significant increase in improvement over the zero-shot baseline, representing a 10\% absolute increase (38\% relative increase). Pass@1 improvements are especially helpful for assisting with software engineering, where it is more helpful to suggest a single correct completion rather than 10 possible completions for the user to select from.

Compared to the gold standards, ILF outperforms both fine-tuning on MBPP gold programs and human-written refinements on the pass@1 metric, yielding 14\% absolute (64\% relative) and 3\% absolute (9\% relative) increases in pass@1 rates, respectively. However, training on human-written refinements yielded comparable pass@10 rates as ILF, which is unsurprising since $\pi_\text{Refine}$ was trained on human-written refinements. When human-written feedback and $\pi_\text{Refine}$-generated refinements are ablated (the ``Ablations'' section of Table \ref{tab:final-results}), ILF also outperforms training on both 1-shot and 2-shot InstructGPT-generated refinements by 17\% and 11\% absolute (89\% and 44\% relative), respectively.

\begin{figure}[ht!]
\begin{center}
    \centerline{\includegraphics[width=\columnwidth]{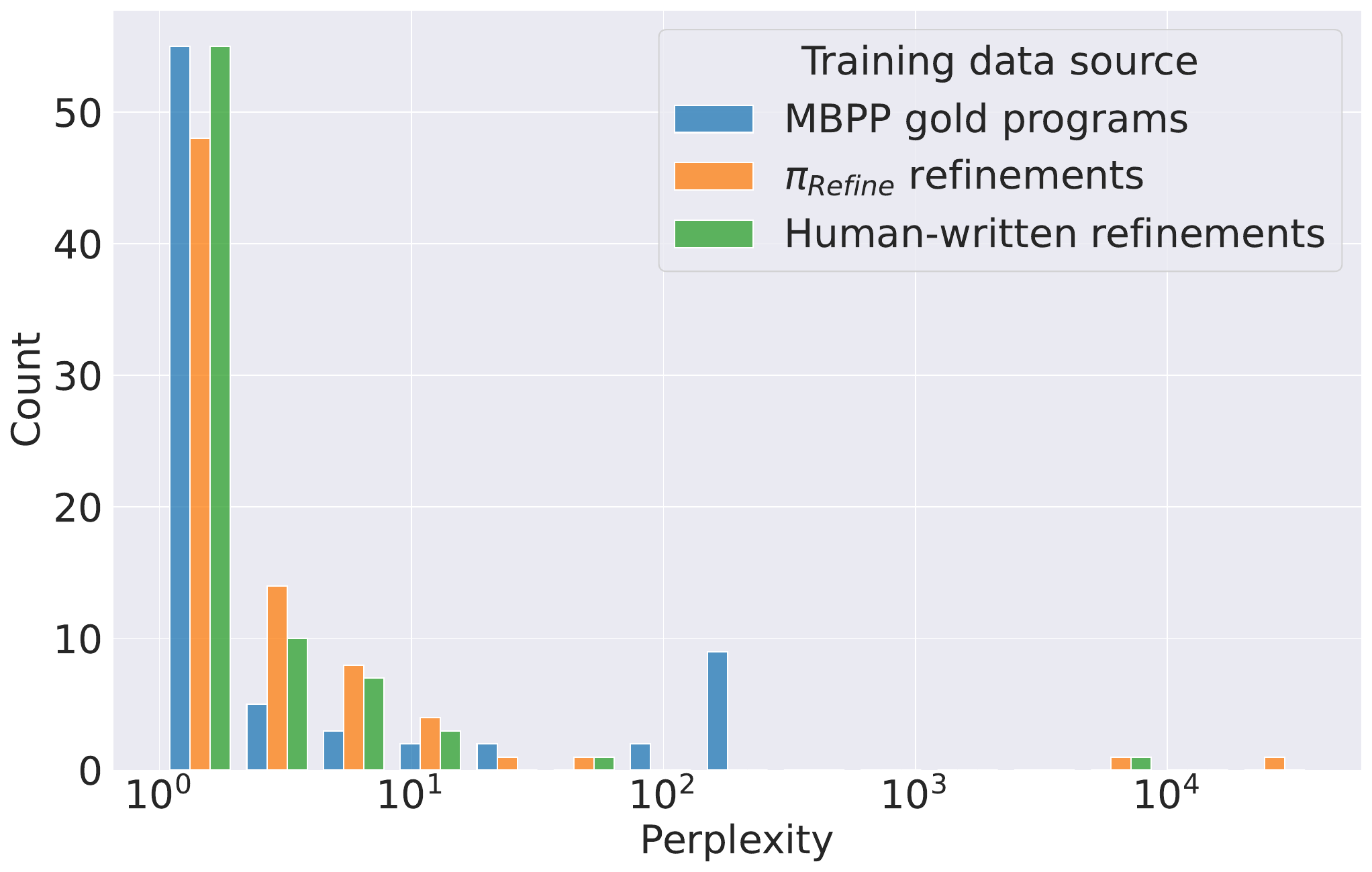}}
    \caption{Histogram of the perplexities of the various training data sources, as measured using a pre-trained \textsc{CodeGen-Mono 6.1B} model.}
    \label{fig:ppl-histogram}
\end{center}
\end{figure}

\begin{figure*}[ht!]
\begin{center}
\centerline{\includegraphics[width=\textwidth]{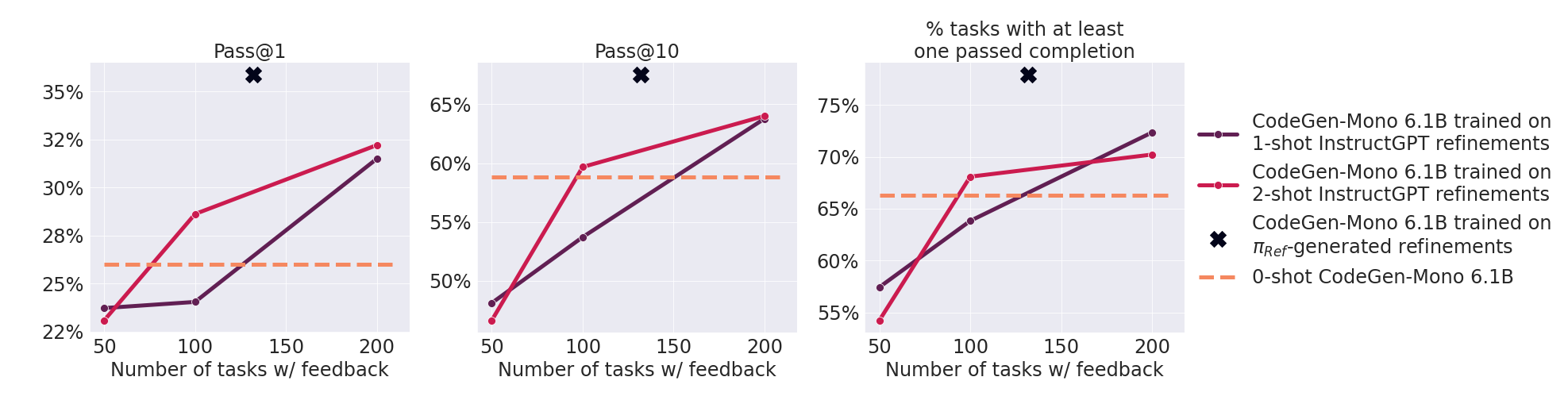}}
\caption{Training dataset size versus \textsc{CodeGen-Mono 6.1B} pass rates on MBPP tasks 11-111 after fine-tuning on InstructGPT-generated refinements, versus the performance of $\pi_{\theta^*}$ (the model produced by our approach). \textit{X} marks the performance of $\pi_{\theta^*}$, whereas the solid lines plot the performance of \textsc{CodeGen-Mono 6.1B} after fine-tuning on correct refinements generated by InstructGPT, using feedback also generated by InstructGPT. The dashed line indicates the zero-shot pass rate of a pre-trained \textsc{CodeGen-Mono 6.1B} model.}
\label{fig:model-feedback-scaling}
\end{center}
\end{figure*}

\paragraph{Analysis of Training Data Sources} However, we also note the surprising fact that merely training on a small sample of the MBPP gold programs did not make a significant difference in accuracy over zero-shot inference. We speculate that the gold programs from the MBPP dataset may be somewhat out-of-distribution for \textsc{CodeGen-Mono} 6.1B. To test this hypothesis, we computed the perplexity of the MBPP gold programs, the $\pi_\text{Refine}$-generated refinements, and the human-written refinements using the pre-trained \textsc{CodeGen-Mono 6.1B} model. The results are shown in Figure \ref{fig:ppl-histogram}. While the distributions of all three data sources look similar, the MBPP dataset contains more high-perplexity programs (\emph{i.e.} programs with perplexity $\geq 10^2$) than either the $\pi_\text{Refine}$-generated refinements or the human-written refinements. As a result, it is likely easier for \textsc{CodeGen-Mono 6.1B} to learn from the latter two datasets, since they are closer to \textsc{CodeGen-Mono 6.1B}'s original distribution while still being functionally correct.

Furthermore, ILF is particularly useful for settings where large amounts of gold code are not available. In this setting, ILF can be thought of as a method of not only generating more training data, but training data that is closer to the model's original outputs in data representation space and that specifically repairs the kinds of bugs that the original model generates. As a result, fine-tuning the model on $\pi_\text{Refine}$-generated refinements does not require adjusting the weights as much as fine-tuning the model on the MBPP gold programs would, even though both training datasets contain the same number of functionally correct programs.



\subsection{Scaling Up Model Feedback Does Not Offer the Same Benefits As Human Feedback}\label{sec:llm-feedback}
Since high quality human feedback can be expensive to collect, we also evaluated how much model feedback might yield the same benefit as our sample of human-written feedback. To do so, we randomly select $k$ tasks from the set of MBPP tasks for which \textsc{CodeGen-Mono 6.1B} did not originally output a correct answer, and prompt InstructGPT to generate both the feedback and the refinement. We then evaluate the refinements for correctness and train \textsc{CodeGen-Mono 6.1B} on the correct refinements. We use $k\in\{50,100,200\}$ and generate 30 output samples at temperature 0.8 for all stages of the experiment. We are limited to these $k$ values due to the small number of tasks we have in MBPP\textsubscript{Train}, but future work may investigate scaling up these experiments by using larger datasets or automatically generating new tasks and unit tests for the training dataset. Further training details are listed in Appendix \ref{appendix:training-details}.

The results are shown in Figure \ref{fig:model-feedback-scaling}. Although increasing the quantity of InstructGPT-generated feedback offers modest improvements in pass rates, these improvements do not yield pass rates as high as those of $\pi_{\theta^*}$, even though $\pi_{\theta^*}$ uses only a total of 122 pieces of feedback throughout its training process (44 for training $\pi_\text{Refine}$ and 78 for generating refinements to train $\pi_{\theta^*}$ on). However, as pre-trained large language models continue to improve dramatically in quality, we expect that this gap between human- and model-written feedback will increasingly narrow.

\begin{table}[th!]
\caption{The proportion of the feedback that addressed each type of bug, for feedback sourced from humans and InstructGPT. Each sample of feedback can be tagged with multiple categories, so the quantities in each column do not necessarily add up to 100\%.}
\label{tab:bug-types}
\begin{small}
\begin{center}
\begin{tabular}{p{4cm}rr}
\toprule
\multirow{2}{=}{Feedback Category} & \multicolumn{2}{c}{\% of Feedback} \\
 & Human & InstructGPT \\
 \midrule
Logic &	30\%  & 46\% \\
Formatting &	36\% & 14\% \\
Missing step &	10\% & 6\%\\
Algebra	 & 10\%  & 8\% \\
Recursion & 4\%  & 14\% \\
Regex	 & 6\%  & 6\% \\
Function semantics	 & 2\% & 4\% \\
Dynamic programming	 & 2\% & 0\% \\
Extra step	 & 0\% & 12\% \\
No feedback needed & 0\% & 14\% \\
Unrelated & 0\% & 8\% \\
\bottomrule
\end{tabular}
\end{center}
\end{small}
\end{table}

\begin{figure}[ht!]
    \centering
    \includegraphics[width=0.8\columnwidth]{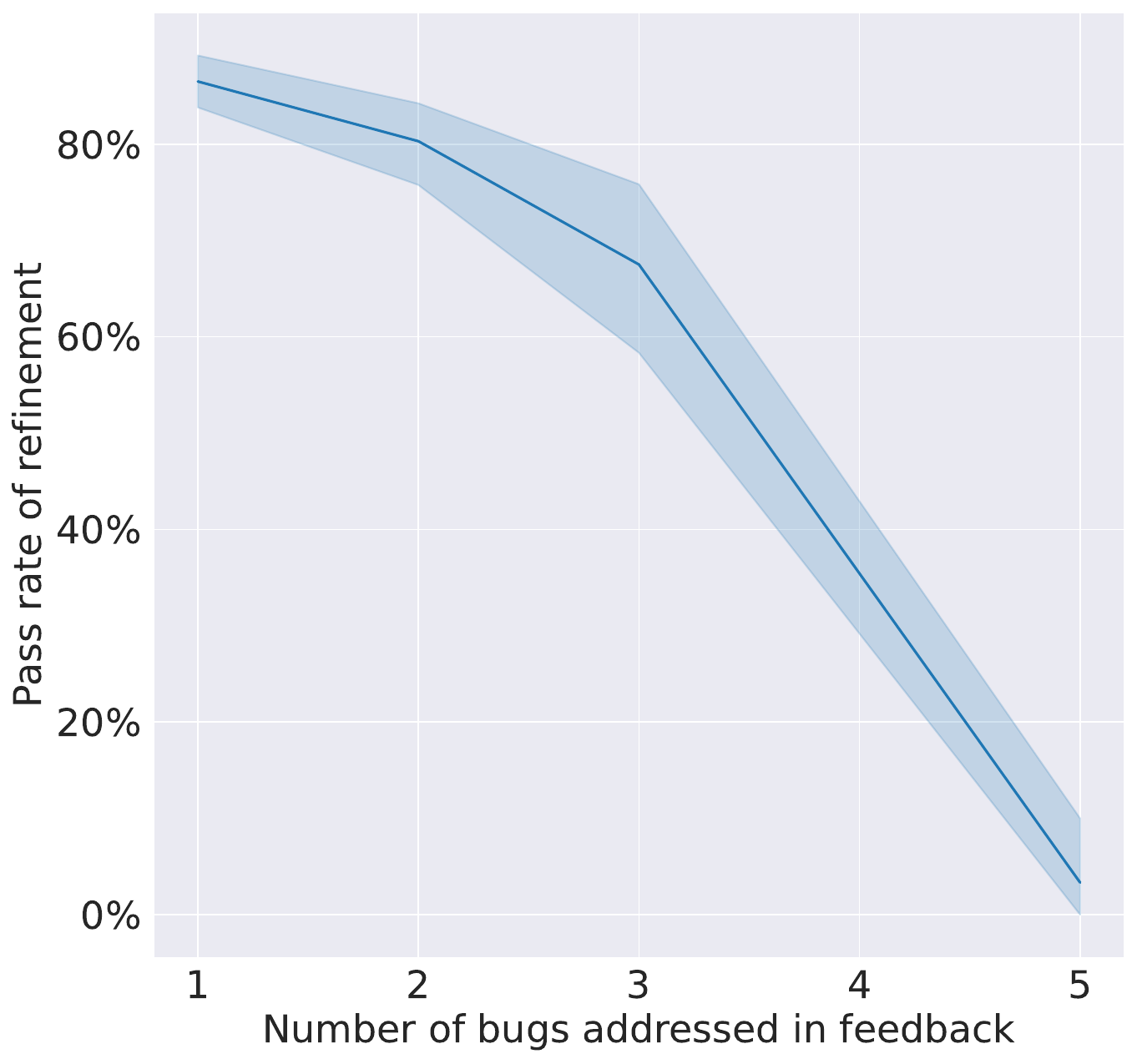}
    \caption{The number of bugs addressed in the feedback versus the pass rate of $\pi_\text{Refine}$'s refinements.}
    \label{fig:num_bugs_vs_pass_rate}
\end{figure}

\begin{table}[th!]
\caption{Descriptive statistics for the human- versus InstructGPT-generated feedback. The * indicates that the metric was computed on the random sample of 50 that we manually inspected, whereas the other metrics are computed from the full dataset.}
\label{tab:feedback-stats}
\begin{small}
\begin{center}
\begin{tabular}{p{3.8cm}R{1.6cm}R{1.6cm}}
\toprule
 & \multicolumn{2}{c}{Source of Feedback} \\
 & Human & InstructGPT \\
 \midrule
Avg. num. of bugs addressed* & 1.8 & 1.1 \\
Avg. num. of words & $68.9 \pm 48.2$ & $24.2 \pm 28.6$\\
\bottomrule
\end{tabular}
\end{center}
\end{small}
\end{table}

\subsection{Human Feedback Is More Informative Than InstructGPT Feedback}\label{sec:human-feedback-more-informative}
To better understand why human feedback produced greater improvements in pass rate than InstructGPT feedback, we randomly selected 50 samples of feedback for each source (\emph{i.e.} human or InstructGPT) and annotated the number and types of bugs that each feedback sample addressed. The results are shown in Tables \ref{tab:bug-types} and \ref{tab:feedback-stats}. We observed that InstructGPT often gave no feedback (\emph{e.g.} ``The code is correct" or ``Great job!"), provided feedback that was irrelevant or incorrect, or restated the task description instead of addressing what should be repaired about the code. Despite this, InstructGPT's refinements were often correct even if the feedback itself wasn't. Human-written feedback addressed more bugs on average and never gave irrelevant feedback. We provide further examples of the differences between human and InstructGPT feedback in Appendix \ref{appendix:human-vs-instructgpt-feedback}.

\subsection{$\pi_\text{Refine}$ Struggles To Incorporate Feedback Addressing Many Bugs}\label{sec:pi-refine-feedback-num-bugs}
Lastly, we explored whether the number of bugs addressed in the feedback affected $\pi_\text{Refine}$'s ability to repair the original code sample. The results are shown in Figure \ref{fig:num_bugs_vs_pass_rate}. The greater the number of bugs addressed, the lower the average pass rate of $\pi_\text{Refine}$'s refinements. This suggests that a promising direction for future work might consist of automatically decomposing the feedback into multiple steps and having $\pi_\text{Refine}$ incorporate the feedback one step at a time. Indeed, \citet{Nijkamp2022CG} show that the \textsc{CodeGen} models are often more effective at following instructions when the instructions are given across multiple turns, and recent Chain-of-Thought work \citep{wei2022chain} illustrates a similar prompting technique.

\section{Related Work}
\paragraph{LLMs for Program Synthesis}
Our work builds on a large body of literature that explores the use of pre-trained LLMs for neural program synthesis. Many general purpose LLMs, although not pre-trained specifically for code generation, have demonstrated impressive proficiency at solving code challenges since they are pre-trained on large corpora of text such as \textsc{The Pile} \citep{gao2020pile} that contain a small percentage of code content \citep{austin2021program, gpt-j, gpt-neox-20b,  Nijkamp2022CG}. Yet other recent LLMs for program synthesis are trained on solely source code files \citep{wang2021codet5, CERT, li2022alphacode, xu2022evaluation}, or on both text and source code documents -- sometimes either in succession \citep{chen2021codex, Nijkamp2022CG, Bai2022TrainingAH}, in a mixed corpus \citep{bigscience2022bloom}, or on mixed natural language-programming language documents \citep{feng-etal-2020-codebert}.

\paragraph{Learning from Human Feedback}
Our algorithm is inspired by a number of past works that have trained models to learn from feedback. A common technique is reinforcement learning from human feedback \citep[RLHF][]{ziegler2019finetuning, stiennon2020learning_to_summarize, ouyang2022instructgpt}, which trains models to satisfy human preferences. However, our algorithm is closer to works that use natural language feedback, rather than comparisons between different choices. \citet{elgohary-etal-2020-speak, austin2021program, Nijkamp2022CG} all demonstrate that code LLM performance generally improves when prompted with natural language feedback, though \citet{Nijkamp2022CG} observes that the feedback is more effective when it is given one step at a time. Our work differs from these in that ILF learns from the feedback at training time, not at inference time.

\citet{Bai2022TrainingAH} also uses natural language feedback during the training process, but as part of an RLHF algorithm instead where the feedback is used to solicit different responses from the digital assistant, the responses are ranked by crowdworkers, and the rankings are used to train the preference model. However, they note that this form of learning from natural language feedback does not measurably improve their code generation model more than simply prompting.

Outside of program synthesis, we show in our other work \citep{scheurer2023training} that ILF is also effective for text summarization. In addition to re-formulating the reward function $R(\cdot)$ for summarization, \citet{scheurer2023training} additionally demonstrates that an instruction-finetuned LLM can evaluate its own outputs and select the best one. Similar to our results on code generation, \citet{scheurer2023training} shows that ILF outperforms all supervised fine-tuning baselines on text summarization. This aligns with numerous other works that have explored supervision via natural language in other ways, such as via explanations \citep[\textit{inter alia}]{camburu2018snli, hase2021can, pruthi2021evaluating, lampinen2022can} and as part of RL systems \citep[\textit{inter alia}]{fidler2017teaching, luketina2019survey, lin2020interactive_rl}.

\section{Conclusion}
We have shown that ILF can significantly improve the quality of a code generation model, even with just a small sample of human-written feedback and refinements. This approach is theoretically justified as minimizing the expected KL divergence between $\pi_\theta$ and a target ground-truth distribution, where we acquire signal from the latter via human-written natural language feedback.

This approach is also appealing because it is not model-specific (in the sense that ILF can be used with any type of base model $\pi_\theta$, assuming the existence of a sufficiently capable LLM to act as $\pi_\text{Refine}$), and can be conducted in multiple rounds to continuously improve the model. Furthermore, it is notable that our approach generates training data that is not only correct, but targets the specific kinds of bugs that the model is likely to output. In essence, it provides an \emph{online} training signal that is missing from the offline pre-training set-up of modern LLMs. Our approach is also remarkably sample-efficient, yielding 38\% and 64\% relative increases in pass@1 rate over the zero-shot baseline and fine-tuning on MBPP data, despite fine-tuning on only 78 examples.


Our work opens up multiple avenues for promising future work. For instance, ILF can be applied iteratively over the course of multiple rounds whenever new information arrives (\emph{e.g.} new Python syntax) or new bugs are discovered. As the pace of progress of modern LLM research continues to accelerate, it may soon be feasible to partially or fully automate the generation of natural language feedback (similar to `RL from AI feedback' \citep[RLAIF;][]{bai2022constitutional} and our experiments in Section \ref{sec:llm-feedback}), greatly reducing both the time and cost necessary for collecting feedback. This direction of work is also particularly appealing because the learning signal is \emph{process-based} rather than outcome-based, which has been shown to mitigate reward hacking and improve the correctness of intermediate reasoning steps \citep{uesato2022solving}. Although further work is required to extend our method, ILF represents an exciting step forward in training LLMs with feedback that is rich, interactive, and sample-efficient.

\section*{Acknowledgements}
We are grateful to Nitarshan Rajkumar, Jason Phang, Nat McAleese, Geoffrey Irving, Jeff Wu, Jan Leike, Cathy Yeh, William Saunders, Jonathan Ward, Daniel Ziegler, Seraphina Nix, Quintin Pope,  Kay Kozaronek, Peter Hase, Talia Ringer, Asa Cooper Stickland, Jacob Pfau, David Lindner,  Lennart Heim, Kath Lumpante, and Pablo Morena for helpful discussions and feedback about the design and implementation of this work. We are additionally thankful to Scott Heiner and Edwin Chen for extensive help with setting up our human annotation workflow and interface. EP thanks the National Science Foundation and Open Philanthropy for fellowship support. JAC is supported by a doctoral grant from the Spanish MECD. AC, SB, and KC are supported by National Science Foundation Awards 1922658 and 2046556. Any opinions, findings, and conclusions or recommendations expressed in this material are those of the author(s) and do not necessarily reflect the views of the National Science Foundation. KC is additionally supported by 42dot, Hyundai Motor Company (under the project Uncertainty in Neural Sequence Modeling) and the Samsung Advanced Institute of Technology (under the project Next Generation Deep Learning: From Pattern Recognition to AI). This project has also benefited from financial support to SB by Eric and Wendy Schmidt (made by recommendation of the Schmidt Futures program), Open Philanthropy, and Apple. We also thank the NYU High-Performance Computing Center for in-kind support and OpenAI for providing access to and credits for their models via the API Academic Access Program.

\bibliography{main}
\bibliographystyle{icml2023}

\newpage
\appendix
\onecolumn
\section{Appendix}

\subsection{Training Details}\label{appendix:training-details}
For the experiments in Section \ref{sec:ilf-results}, we run a hyperparameter sweep for all methods except for ILF. The hyperparameter value ranges that we sweep include learning rate $\in\{1.0^{-6}, 5.0^{-6}, 1.0^{-5}\}$, batch size $\in\{32, 64, 128\}$, and number of epochs $\in \{1, 2, 5\}$. The tasks for the training and validation datasets are from MBPP\textsubscript{Train} and MBPP\textsubscript{Refine}, respectively, while the programs are sourced from the method (\emph{e.g.} InstructGPT, MBPP, human-written, or zero-shot \textsc{CodeGen-Mono} 6.1B). For ILF, we use the best hyperparameters obtained for the sweep over MBPP programs instead of sweeping over ILF-generated programs, since the tasks in MBPP\textsubscript{Refine} are already used to train $\pi_\text{Refine}$. All pass rates reported in Table \ref{tab:final-results} are obtained by evaluating each method on MBPP\textsubscript{Test} using the best hyperparameters found during the sweep on MBPP\textsubscript{Refine}.

For the experiments in Section \ref{sec:llm-feedback}, we separately tune hyperparameters for each size of dataset. As in our other experiments, we train and validate using the tasks from MBPP\textsubscript{Train} and MBPP\textsubscript{Refine}, respectively, coupled with the refinements generated by InstructGPT that pass the unit test suites. We sweep the same hyperparameter value ranges as the experiments in the previous section (\emph{i.e.} learning rate $\in\{1.0^{-6}, 5.0^{-6}, 1.0^{-5}\}$, batch size $\in\{32, 64, 128\}$, and number of epochs $\in \{1, 2, 5\}$).

We implement all experimental pipelines with the HuggingFace transformers (v4.12.5) \citep{wolf-etal-2020-transformers}, Huggingface datasets (v2.7.1) \citep{lhoest-etal-2021-datasets}, and Pytorch (v1.11) \citep{Paszke_PyTorch_An_Imperative_2019} libraries.

\newpage
\subsection{Annotator Instructions}\label{appendix:annotator-instructions}
\begin{figure*}[th!]
    \centering
    \includegraphics[width=\textwidth]{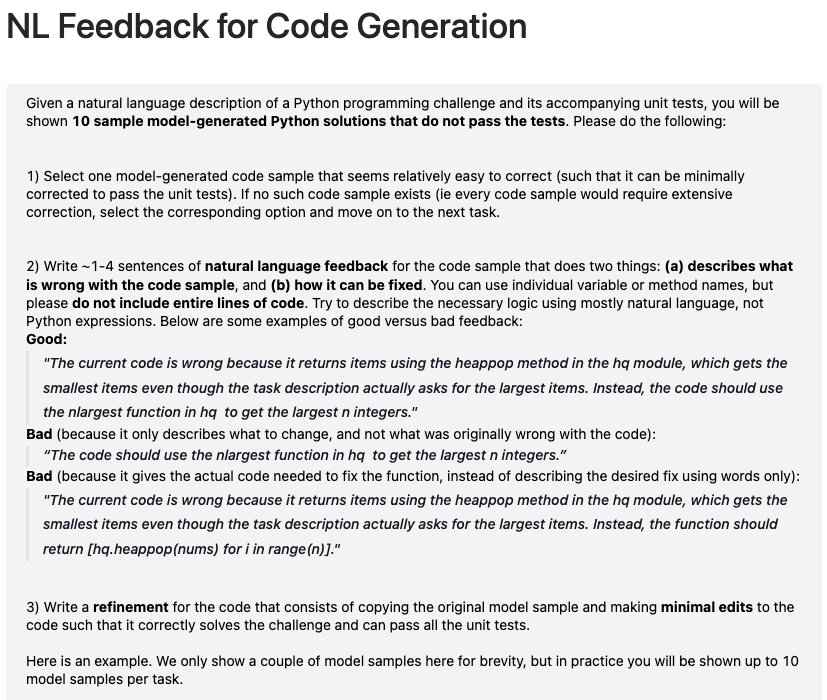}
    \caption{An example of the instructions given to Surge annotators.}
\end{figure*}

\newpage
\subsection{Examples of Human Versus InstructGPT Feedback}\label{appendix:human-vs-instructgpt-feedback}
{\setlength{\extrarowheight}{5pt}%
\begin{longtable}[th!]{p{0.4\textwidth}p{0.25\textwidth}p{0.25\textwidth}}

    \caption{Examples of natural language feedback written by humans versus InstructGPT for incorrect programs generated by \textsc{CodeGen-Mono 61.B}.}
    \label{tab:human-vs-instructgpt-feedback} \\
    \toprule
    Incorrect program generated by \textsc{CodeGen-Mono 6.1B} & Human feedback & InstructGPT feedback \\ \midrule
    \endfirsthead

    \caption*{\tablename\ \thetable{}: Continued from previous page} \\
    \toprule
    Incorrect program generated by \textsc{CodeGen-Mono 6.1B} & Human feedback & InstructGPT feedback \\ \midrule
    \endhead
       \vspace{-10pt}
\begin{lstlisting}[language=Python]
"""
Write a function to assign frequency to each tuple in the given tuple list.

>>> Example: assign_freq([(6, 5, 8), (2, 7), (6, 5, 8), (6, 5, 8), (9, ), (2, 7)] ) = '[(6, 5, 8, 3), (2, 7, 2), (9, 1)]'
"""

from collections import Counter 
def assign_freq(test_list):
        """ 
        >>> assign_freq([(6, 5, 8), (2, 7), (6, 5, 8), (6, 5, 8), (9, ), (2, 7)] ) = '[(6, 5, 8, 3), (2, 7, 2), (9, 1)]' 
        """
        counter = Counter(x for x in test_list)
        res = []
        for key, val in counter.items():
                temp = (key, val)
                res.append(temp)
        return str(res) 
\end{lstlisting} & ``The only issue with this code is that it does not dereference the key. When you set temp = (key, val), the key is an entire tuple instead of the values in the tuple. Thus, you need the asterisks to get the values" &  ``The function works as intended." \\
       \vspace{-10pt}
\begin{lstlisting}[language=Python]
"""
Write a function to get the angle of a complex number.

>>> Example: angle_complex(0,1j)=1.5707963267948966 
"""

import cmath
def angle_complex(a,b):
        return cmath.phase(complex(a,b))
\end{lstlisting} & ``Only one thing is wrong in the code that complex method can take the complex number as argument example complex(a+b) not like complex(a,b) we have to replace, by + and the code is fine." &  ``The code should use the atan2 function to get the angle of a complex number." \\
       \vspace{-10pt}
\begin{lstlisting}[language=Python]
"""
Write a function to count the same pair in three given lists.

>>> Example: count_samepair([1,2,3,4,5,6,7,8],
    [2,2,3,1,2,6,7,9],
    [2,1,3,1,2,6,7,9])=3
"""

def count_samepair(list1,list2,list3):
    count = 0
    for i in range(len(list1)):
        for j in range(len(list2)):
            if list1[i]==list2[j] and list2[j]==list3[i]:
                count+=1
    return count
\end{lstlisting} &  ``One extra for loop for list2 is causing the issue, it finds more matching elements between list1,2,3. The possible solution for the code is to remove the unnecessary for loop of list2 iteration, and check all the lists same index element at the time, and return the count of this element." & ``The code above is very inefficient. It is checking every possible combination of pairs of elements in the three lists. It would be much more efficient to use a set to keep track of the pairs that have already been seen." \\
       \vspace{-10pt}
\begin{lstlisting}[language=Python]
"""
Write a function to find number of lists present in the given tuple.

>>> Example: find_lists(([1, 2, 3, 4], [5, 6, 7, 8])) = 2
"""

def find_lists(Input):
    result = 0
    for x in Input:
        if isinstance(x, list):
            result += 1
    return result
\end{lstlisting} &  ``The code works for tuples of more than one list, but not in case of one single list. Python goes one level down and reads the elements when given a single list and does not see an instance of list. I would just check for the case of `type(Input) == list' and turn the input into a 1-sized list of lists in this case." & ``The function find\_lists needs to return the number of lists present in the given tuple, not the number of items in the tuple." \\
       \vspace{-10pt}
\begin{lstlisting}[language=Python]
"""
Write a function to calculate magic square.

>>> Example: magic_square_test([[7, 12, 1, 14], [2, 13, 8, 11], [16, 3, 10, 5], [9, 6, 15, 4]])=True
"""

def magic_square_test(my_matrix):
    """
    >>> magic_square_test([[7, 12, 1, 14], [2, 13, 8, 11], [16, 3, 10, 5], [9, 6, 15, 4]])=True
    """
    def is_square(x):
        return x==x*x
    def is_magic_square(my_matrix):
        sides=[]
        for i in range(len(my_matrix)):
            sides.append(sum(my_matrix[i]))
        if not is_square(sum(sides)):
            return False
        sides=[]
        for i in range(len(my_matrix)):
            sides.append(sum([my_matrix[i][j] for j in range(len(my_matrix))]))
        if not is_square(sum(sides)):
            return False
        diagonals=[]
        for i in range(len(my_matrix)):
            diagonals.append(my_matrix[i][i])
        if not is_square(sum(diagonals)):
            return False
        diagonals=[]
        for i in range(len(my_matrix)):
            diagonals.append(my_matrix[i][len(my_matrix)-1-i])
        if not is_square(sum(diagonals)):
            return False
        return True
    return is_magic_square(my_matrix)
\end{lstlisting} & ``The logic behind checking whether the sum of rows, columns, and diagonals is a magic number or not is not correct, the code checks if the sum of squares of itself in the is\_square method is wrong. Instead of checking all the sums of rows, columns, and diagonals at every point, the logic should be like saving all the sums for rows, columns, and diagonals in one array(sides) and then checking if there are all same values present in this array because for magic\_square\_test all the sums should be equal." & ``The function magic\_square\_test should return whether my\_matrix is a magic square." \\
       \vspace{-10pt}
\begin{lstlisting}[language=Python]
"""
Write a python function to count characters at same position in a given string (lower and uppercase characters) as in english alphabet.

>>> Example: count_char_position("xbcefg") = 2
"""

def count_char_position(str1):
    counter = 0
    for i in range(0, len(str1)):
        if str1[i] == str1[i].upper():
            counter += 1
    return counter
\end{lstlisting} & ``a) The first thing that the code sample gets incorrect is that it increments the counter when each character in the string is an upper case character. The prompt is asking for how many characters in str1 have the same numerical position as their respective alphabetic counterpart and the code does not check for this at all. b) The way to fix it would be to create a separate j variable that keeps track of the ordinal value that corresponds with the alphabetic character that we would expect to be compared to the respective character in str1. j and i would increment at the same time so that the code is both iterating through the alphabet AND iterating through str1 simultaneously. The other fix would be to automatically convert str1 to lowercase so that we don't have to worry about checking the ordinal codes of both the lowercase and uppercase versions of the same alphabetic character." & ``Great job!" \\
    \hline
    
    \label{tab:my_label}
\end{longtable}

\end{document}